\documentclass[12pt,fleqn]{article}
\usepackage{amssymb}
\newtheorem{thm}{Theorem}[section]

\def\ha{\mbox{$\cal H$}}
\def\la{\langle}
\def\ra{\rangle}

\def\ie{\emph{i.e.}}
\def\eg{\emph{e.g.}}
\def\eq{\begin{equation}}
\def\en{\end{equation}}
\def\ot{\otimes}
\def\id{\mbox{\small id}}
\def\no{\mbox{\footnotesize \#}}

\def\Z{\mathbb{Z}}
\def\C{\mathbb{C}}
\def\Re{\mathbb{R}}
\def\R{{\cal R}}
\begin{document}
\begin{titlepage}
\rightline{CRM-2507}
\rightline{PUPT-1731}
\rightline{LMU-TPW 97-24}
\vspace{2em}
\begin{center}{\bf\Large AKS scheme for face and 
Calogero-Moser-Sutherland type models}\\[2em]
Branislav Jur\v co${}^{*}$ and Peter Schupp${}^{**}$\\[2em]
{\sl ${}^*$CRM, Universit\'e de Montr\'eal\\Montr\'eal (Qc), 
H3C 3J7, Canada}\\[2em]
{\sl ${}^{**}$Department of Physics,
Princeton University\\Princeton, NJ 08544-0708, USA} \\[6em]
\end{center}
\begin{abstract}
We give the construction of quantum Lax equations for IRF models 
and difference versions of Calogero-Moser-Sutherland
models introduced by Ruijsenaars. We solve the equations using
factorization properties of the underlying face Hopf 
algebras/elliptic quantum groups. 
This construction is in the spirit of the
Adler-Kostant-Symes method and generalizes our previous work 
to the case of face
Hopf algebras/elliptic quantum groups with dynamical
R-matrices.
\end{abstract}

\vfill
\noindent \hrule
\vskip.2cm
\hbox{{\small{\it e-mail: }}{\small \quad jurco@CRM.UMontreal.CA,\quad
schupp@princeton.edu}}
\end{titlepage}
\newpage
\setcounter{page}{1}

\section{Introduction}

Face Hopf algebras \cite{Hay} have been found to be the
algebraic structure that 
underlies some particularly interesting
integrable models of statistical physics. They generalize Hopf algebras
and quantum groups. Another closely related
generalization of quantum groups, the
so-called elliptic quantum groups, were introduced by Felder \cite{Fel}
in the context of IRF (face) models
\cite{Bax}. 
Face Hopf algebras and elliptic quantum groups
play the same role for face models as quantum groups do for vertex models.
The R-matrices are now replaced by dynamical R-matrices,
which first appeared in the context of Liouville string field theory
\cite{Ger}; they can
be understood as a reformulation of the Boltzmann weights in
Baxters solutions of the face-type Yang-Baxter equation.
A partial classification of the dynamical R-matrices is given in 
\cite{EV}.
Recently another type of integrable quantum systems, known as Ruijsenaars
models \cite{Rui}, 
and their various limiting cases have been shown to be connected to
quantum groups and elliptic quantum groups
\cite{Has,FV,ABB,AF,EK}
through different approaches.

The Calogero-Moser-Sutherland class of integrable 
models describe the motion of
particles on a one-dimensional line or circle interacting
via pairwise potentials that are given by 
Weierstrass elliptic functions and their various degenerations.
The simplest case is an inverse $r^2$ potential.
The Ruijsenaars-Schneider model is a relativistic generalization, whose
quantum mechanical version, the Ruijsenaars model, is the model that
we are interested in here.

The Hamiltonian of the Ruijsenaars model 
for two particles with coordinates $x_1$ and $x_2$ has the form
\[
\ha = \left\{\frac{\theta(\frac{c \eta}{2} - \lambda)}{\theta(-\lambda)}
t^{(\lambda)}_1
+ \frac{\theta(\frac{c \eta}{2} + \lambda)}{\theta(\lambda)} 
t^{(\lambda)}_2\right\} , 
\]
where $\lambda = x_1 - x_2$. Here $c \in \C$ is the coupling 
constant, $\eta$ is the
relativistic deformation parameter, the $\theta$-function is 
given in (\ref{theta}); 
we have set $\hbar =1$. The Hamiltonian acts  on a wave function as
\[
\ha \,\psi(\lambda) = 
\frac{\theta(\frac{c \eta}{2} 
- \lambda)}{\theta(-\lambda)}\psi(\lambda - \eta) 
+ \frac{\theta(\frac{c \eta}{2} 
+ \lambda)}{\theta(\lambda)}\psi(\lambda + \eta),
\]
the $t_i^{(\lambda)}$ that appear in the Hamiltonian are hence
shift-operators in the variable 
$\lambda$; in the present case of
two particles they generate a one-dimensional
graph:
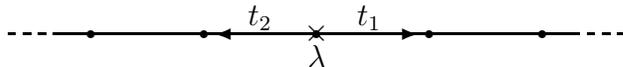
\begin{center}
\unitlength 0.50mm
\linethickness{0.4pt}
\thicklines
\begin{picture}(164.00,20.00)
\put(40.00,10.00){\circle*{2.00}}
\put(100.00,10.00){\circle*{2.00}}
\put(73.00,10.00){\vector(1,0){24.00}}
\put(67.00,10.00){\vector(-1,0){24.00}}
\put(70.00,7.00){\makebox(0,0)[ct]{$\lambda$}}
\put(84.00,11.00){\makebox(0,0)[cb]{$t_1$}}
\put(55.00,11.00){\makebox(0,0)[cb]{$t_2$}}
\put(10.00,10.00){\circle*{2.00}}
\put(130.00,10.00){\circle*{2.00}}
\put(70.00,10.00){\circle*{2.00}}
\put(70.00,10.00){\makebox(0,0)[cc]{$\times$}}
\thinlines
\put(0.00,10.00){\line(1,0){140.00}}
\put(142.00,10.00){\line(1,0){2.00}}
\put(146.00,10.00){\line(1,0){2.00}}
\put(150.00,10.00){\line(1,0){2.00}}
\put(-12.00,10.00){\line(1,0){2.00}}
\put(-8.00,10.00){\line(1,0){2.00}}
\put(-4.00,10.00){\line(1,0){2.00}}
\end{picture}
\end{center}
Relative to a fixed vertex $\lambda$ the
vertices of this graph 
are at points $\eta\cdot\Z \in \Re$, the (ordered) paths 
connect neighboring vertices and are hence
intervals in $\Re$. 
There are two paths per interval,
one in positive, one in negative direction.

Now consider a $N$-particle Ruijsenaars model: 
The graph relative to a fixed $\lambda$ is
then a $N -1$ dimensional hyper-cubic lattice. 
Relative to
$\lambda \in \Re^{N-1}$ the vertices are at points 
$\left(\eta \cdot \Z\right)^{\times (N-1)}$.
This picture can obviously be generalized and in the following we
shall consider an arbitrary ordered graph.

\paragraph{\it Remark:} For the Ruijsenaars system 
$\lambda$ can take any value in 
$\Re^{N-1}$, so we
are a priori let to a huge graph that 
consists of a continuous
family of disconnected graphs. Since the graphs are 
disconnected it is
sufficient to consider one.
Later we will in fact 
write all expressions with respect to one fixed vertex 
$\lambda$. This
will lead to equations containing explicit 
shifts on the graph and thus to dynamical $R$-matrices.
Apart from the physical interpretation there is a priori no reason
to restrict the values of particle-coordinates  at the level 
of the shift operator to $\Re$. We can and shall everywhere
in the following take $\C$ instead of $\Re$. 

\subsection{Face algebras}
 
The Hilbert space of the model 
will be build from vector spaces on paths of fixed length.
As we shall argue in the following, the operators
on this Hilbert space can be chosen to be elements of
a Face Algebra $F$ \cite{Hay} (or weak $C^*$-Hopf
algebras \cite{Vol,Boe}, which is essentially the same with a $*$-structure).

There are two commuting projection operators $e^i, e_i \in F$ 
for each
vertex of the graph (projectors 
onto bra's and ket's corresponding to vertex $i$):
\eq
e_i e_j = \delta_{ij} e_i, \quad e^i e^j = \delta_{ij} e^i, \quad
{\textstyle\sum} e_i = {\textstyle\sum} e^i = 1 .
\en
$F$ shall  be equipped with a coalgebra structure such that the
combination $e^i_j \equiv e^i e_j = e_j e^i$ is a
corepresentation:
\eq
\Delta(e^i_j) = {\textstyle\sum}_k e^i_k \ot e^k_j,
\quad \epsilon(e^i_j) = \delta_{ij}.
\en
It follows that $\Delta(1) = \sum_k e_k \ot e^k \neq 1 \ot 1$ 
(unless the graph has only a single vertex) 
-- this is a key feature of face algebras (and weak 
$C^*$-Hopf
algebras).

So far we have considered matrices with indices that are vertices,
\ie\ paths of length zero. In the given setting it is natural to also
allow paths of fixed length on a finite
oriented graph $\cal G$
as matrix indices. To illustrate this, here 
are some paths of length 3 on a graph that is
a square lattice, \eg\ on part of a graph corresponding to a 
particular Ruijsenaars system:
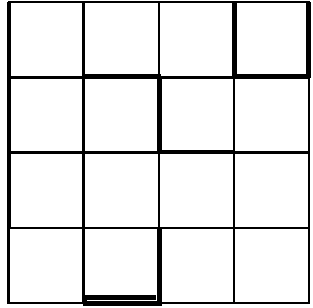
\begin{center}
\unitlength 1.00mm
\linethickness{0.4pt}
\begin{picture}(40.00,40.00)
\thinlines
\put(0.00,0.00){\line(0,1){40.00}}
\put(0.00,40.00){\line(1,0){40.00}}
\put(40.00,40.00){\line(0,-1){40.00}}
\put(40.00,0.00){\line(-1,0){40.00}}
\put(0.00,10.00){\line(1,0){40.00}}
\put(40.00,20.00){\line(-1,0){40.00}}
\put(0.00,30.00){\line(1,0){40.00}}
\put(10.00,40.00){\line(0,-1){40.00}}
\put(20.00,40.00){\line(0,-1){40.00}}
\put(30.00,40.00){\line(0,-1){40.00}}
\thicklines
\put(0.00,40.00){\line(0,-1){30.00}}
\put(10.00,0.00){\line(1,0){10.00}}
\put(20.00,-0.30){\line(0,1){10.30}}
\put(10.00,30.00){\line(1,0){10.30}}
\put(20.00,30.00){\line(0,-1){10.00}}
\put(20.00,20.00){\line(1,0){10.00}}
\put(30.00,40.00){\line(0,-1){10.00}}
\put(30.00,30.00){\line(1,0){10.00}}
\put(40.00,30.00){\line(0,1){10.00}}
\put(10.00,0.00){\line(0,1){0.50}}
\put(10.00,0.60){\line(1,0){9.60}}
\put(0.20,40.00){\line(0,-1){30.00}}
\put(10.00,-0.20){\line(1,0){10.00}}
\put(20.20,-0.30){\line(0,1){10.30}}
\put(10.00,30.20){\line(1,0){10.30}}
\put(20.20,30.00){\line(0,-1){10.00}}
\put(20.00,20.20){\line(1,0){10.00}}
\put(30.20,40.00){\line(0,-1){10.00}}
\put(30.00,30.20){\line(1,0){10.00}}
\put(39.80,30.00){\line(0,1){10.00}}
\put(10.20,-0.30){\line(0,1){1.20}}
\put(10.00,0.80){\line(1,0){9.60}}
\end{picture}
\end{center}
We shall use capital letters 
to label paths. A path $P$ has an origin (source) $\cdot P$,
an end (range) $P \cdot$ and a length 
$\no P$. 
%
%
Two paths $Q$, $P$ can be concatenated to form a new path
$Q \cdot P$, if the end of the first path coincides with the start of
the second path, \ie\ if 
$Q \cdot = \cdot P$ (this explains our choice
of notation).

The important point is, that the symbols
$T^A_B$, where $\no A = \no B \geq 0$,
with relations
\eq
\Delta\left( T^A_B \right) = \sum_{A'} T^A_{A'} \ot T^{A'}_B  \label{delta}
\quad (\no A = \no A' = \no B) 
\en
\eq
\epsilon( T^A_B ) = \delta_{A B} \label{eps}
\en
\eq
T^A_B T^C_D = \delta_{A\cdot,\cdot C} \delta_{B\cdot,\cdot D} T^{A \cdot
C}_{B \cdot D} \label{tt}
\en
span an object that obeys the axioms of a face algebra.
Relations (\ref{delta}) and (\ref{eps}) make $T^A_B$ a
corepresentation; (\ref{tt}) is the rule for combining
representations.
The axioms of a face algebra can be found in \cite{Hay,Hay1}.
\paragraph{\it Pictorial representation:}
$$T^A_B \,\sim\;
\unitlength1mm
\begin{picture}(10,10)(0,4)\small
\put(0,0){\line(1,0){10}}
\put(0,10){\line(1,0){10}}
\multiput(0,0)(0,1){10}{\line(0,1){0.5}}
\multiput(10,0)(0,1){10}{\line(0,1){0.5}}
\put(0,5){\vector(0,1){1}}
\put(10,5){\vector(0,1){1}}
\put(5,0){\vector(1,0){1}}
\put(5,10){\vector(1,0){1}}
\put(5,-1){\makebox(0,0)[ct]{$B$}}
\put(5,11){\makebox(0,0)[cb]{$A$}}
\end{picture}
\quad
T^A_B T^C_D \,\sim\;
\begin{picture}(20,10)(0,4)\small
\put(0,0){\line(1,0){10}}
\put(0,10){\line(1,0){10}}
\multiput(0,0)(0,1){10}{\line(0,1){0.5}}
\multiput(10,0)(0,1){10}{\line(0,1){0.5}}
\put(0,5){\vector(0,1){1}}
\put(10,5){\vector(0,1){1}}
\put(5,0){\vector(1,0){1}}
\put(5,10){\vector(1,0){1}}
\put(5,-1){\makebox(0,0)[ct]{$B$}}
\put(5,11){\makebox(0,0)[cb]{$A$}}
\put(10,0){\line(1,0){10}}
\put(10,10){\line(1,0){10}}
\multiput(20,0)(0,1){10}{\line(0,1){0.5}}
\put(20,5){\vector(0,1){1}}
\put(15,0){\vector(1,0){1}}
\put(15,10){\vector(1,0){1}}
\put(15,-1){\makebox(0,0)[ct]{$D$}}
\put(15,11){\makebox(0,0)[cb]{$C$}}
\end{picture}\quad
\Delta T^A_B = \sum_{A'} T^A_{A'} \otimes T^{A'}_B \,\sim\;
\begin{picture}(15,10)(0,9)\small
\put(0,20){\line(1,0){10}}
\multiput(0,10)(0,1){10}{\line(0,1){0.5}}
\multiput(10,10)(0,1){10}{\line(0,1){0.5}}
\put(0,15){\vector(0,1){1}}
\put(10,15){\vector(0,1){1}}
\put(5,20){\vector(1,0){1}}
\put(5,21){\makebox(0,0)[cb]{$A$}}
\put(0,0){\line(1,0){10}}
\put(0,10){\line(1,0){10}}
\multiput(0,0)(0,1){10}{\line(0,1){0.5}}
\multiput(10,0)(0,1){10}{\line(0,1){0.5}}
\put(0,5){\vector(0,1){1}}
\put(10,5){\vector(0,1){1}}
\put(5,0){\vector(1,0){1}}
\put(5,10){\vector(1,0){1}}
\put(5,-1){\makebox(0,0)[ct]{$B$}}
\end{picture}\vspace{5mm}
$$
The dashed paths indicate the $F$-space(s), their orientation is from lower to
upper ``index''. Inner paths are summed over.

Often it is convenient not to consider a particular representation, \ie\
$T$-matrices corresponding to paths of a given fixed length but rather an
abstract, universal $T$.
Heuristically one can think of the universal $T$ as an abstract matrix or
group element, but it does in fact simply provide an alternative to the
usual Hopf algebra notation:
$T$, $T_1 \ot T_1$, $T_1 T_2$, \eg\ correspond to the identity map,
coproduct map and 
multiplication map of $F$ respectively. (More details are given in the
appendix.)
Keeping this in mind we shall---without loss of generality---nethertheless 
use a formal
notation that treats $T$ as if it was the
canonical element $T_{12}$ of $U \ot F$, 
where $U$ is the dual of $F$ via the
pairing $\la \: , \: \ra$.\footnote{Here 
and in the following we 
shall frequently suppress the
the second index of $T$; it corresponds to the $F$-space.
The displayed expressions are
short for $\la T_{12},f\ot \id\ra = f \in F$,
$\la T_{12} T_{13}, f \ot \id \ot \id\ra 
= \Delta\, f \in F \ot F$
and $\la T_{13} T_{23}, f\ot g\ot \id\ra = f g \in F$.}
\[ 
\la T , f \ra = f, \quad \la T_1 \ot T_1, f \ra = \Delta\,f, \quad
\la T_1 T_2 , f \ot g \ra = f g ; \quad f , g \in F
\]

A face \emph{Hopf} algebra has an anti-algebra and anti-coalgebra endomorphism,
called the antipode and denoted by $S$ or---in the universal tensor
formalism---by $\tilde T$: $\la \tilde T, f\ra = S(f)$. The antipode
satisfies some compatibility conditions with the
coproduct that are given in the appendix.
\paragraph{\it Remark:}
In the limit of a graph with a single vertex a face Hopf algebra
is the same as a Hopf algebra. Ordinary matrix indices correspond
to closed loops in that case.

\subsection{Boltzmann weights}

By dualization we can describe a
coquasitriangular structure of $F$ by giving
a quasitriangular structure  for $U$.
The axioms \cite{Hay} for a quasitriangular face algebra are similar to
those of a quasitriangular Hopf algebra; there is a universal
$R \in U \ot U$ that controls the non-cocommutativity of the coproduct in $U$
and the non-commutativity of the
product in $F$,
\eq
R T_1 T_2 = T_2 T_1 R , \quad \tilde R T_2 T_1 = T_1 T_2 \tilde R,
\quad \tilde R \equiv (S \ot \id)(R),
\label{rtt}
\en
however the antipode of $R$ is
no longer inverse of $R$ but rather 
\eq 
\tilde R R = \Delta(1) , \quad R \tilde R = \Delta'(1) .
\en
The numerical 
``$R$-matrix'' obtained by contracting $R$ with two face corepresentations
is given by the face Boltzmann weight $W$:
\[ 
\la \R , T^A_B \ot T^C_D \ra \; = \; R^{A C}_{B D}  
\;\equiv\; W\Big( {C {B \atop A} D} \Big)
\quad\sim\qquad
\unitlength1mm
\begin{picture}(10,7)(-2,4)\small
\put(0,0){\line(1,0){10}}
\put(0,0){\line(0,1){10}}
\put(10,10){\line(0,-1){10}}
\put(10,10){\line(-1,0){10}}
\put(0,5){\vector(0,-1){1}}
\put(10,5){\vector(0,-1){1}}
\put(5,0){\vector(1,0){1}}
\put(5,10){\vector(1,0){1}}
\put(5,-1){\makebox(0,0)[ct]{$A$}}
\put(5,11){\makebox(0,0)[cb]{$B$}}
\put(-1,5){\makebox(0,0)[rc]{$C$}}
\put(11,5){\makebox(0,0)[lc]{$D$}}
\end{picture}\]\vspace{4mm}
The pictorial representation makes sense since 
the Boltzmann weight is zero unless $C\cdot A$ and $B\cdot D$
are valid paths with common source and range
as will be discussed in more detail
below.
Also note that $T^A_B \mapsto \la \R , T^A_B \ot T^C_D \ra$ is a
\emph{representation} of the matrix elements of $(T^A_B)$ while
$T^C_D \mapsto \la \R , T^A_B \ot T^C_D \ra$ is an \emph{anti-representation}.
Consistent with our 
pictorial representation for the $T$-matrices
we see that the orientation of the
paths in $F$-space remain the same for the first case but are reversed for the
latter:
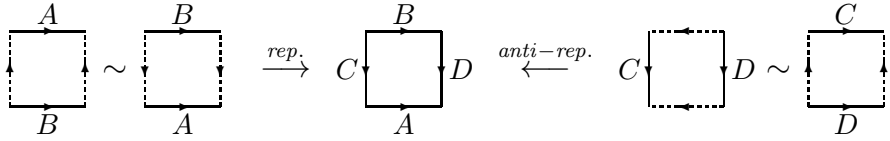
\[\unitlength1mm
\begin{picture}(10,8)(0,4)\small
\put(0,0){\line(1,0){10}}
\put(0,10){\line(1,0){10}}
\multiput(0,0)(0,1){10}{\line(0,1){0.5}}
\multiput(10,0)(0,1){10}{\line(0,1){0.5}}
\put(0,5){\vector(0,1){1}}
\put(10,5){\vector(0,1){1}}
\put(5,0){\vector(1,0){1}}
\put(5,10){\vector(1,0){1}}
\put(5,-1){\makebox(0,0)[ct]{$B$}}
\put(5,11){\makebox(0,0)[cb]{$A$}}
\end{picture}\; \sim \;
\begin{picture}(10,10)(0,4)\small
\put(0,0){\line(1,0){10}}
\put(0,10){\line(1,0){10}}
\multiput(0,0)(0,1){10}{\line(0,1){0.5}}
\multiput(10,0)(0,1){10}{\line(0,1){0.5}}
\put(0,5){\vector(0,-1){1}}
\put(10,5){\vector(0,-1){1}}
\put(5,0){\vector(1,0){1}}
\put(5,10){\vector(1,0){1}}
\put(5,-1){\makebox(0,0)[ct]{$A$}}
\put(5,11){\makebox(0,0)[cb]{$B$}}
\end{picture}\quad
\stackrel{\it rep.}{\longrightarrow}\quad
\begin{picture}(10,10)(-2,4)\small
\put(0,0){\line(1,0){10}}
\put(0,0){\line(0,1){10}}
\put(10,10){\line(0,-1){10}}
\put(10,10){\line(-1,0){10}}
\put(0,5){\vector(0,-1){1}}
\put(10,5){\vector(0,-1){1}}
\put(5,0){\vector(1,0){1}}
\put(5,10){\vector(1,0){1}}
\put(5,-1){\makebox(0,0)[ct]{$A$}}
\put(5,11){\makebox(0,0)[cb]{$B$}}
\put(-1,5){\makebox(0,0)[rc]{$C$}}
\put(11,5){\makebox(0,0)[lc]{$D$}}
\end{picture}\qquad
\stackrel{\it anti-rep.}{\longleftarrow}\quad
\begin{picture}(10,10)(-2,4)\small
\put(0,0){\line(0,1){10}}
\put(10,10){\line(0,-1){10}}
\multiput(10,10)(-1,0){10}{\line(-1,0){0.5}}
\multiput(10,0)(-1,0){10}{\line(-1,0){0.5}}
\put(0,5){\vector(0,-1){1}}
\put(10,5){\vector(0,-1){1}}
\put(5,0){\vector(-1,0){1}}
\put(5,10){\vector(-1,0){1}}
\put(-1,5){\makebox(0,0)[rc]{$C$}}
\put(11,5){\makebox(0,0)[lc]{$D$}}
\end{picture}
\quad\;\; \sim \;
\begin{picture}(10,10)(0,4)\small
\put(0,0){\line(1,0){10}}
\put(0,10){\line(1,0){10}}
\multiput(0,0)(0,1){10}{\line(0,1){0.5}}
\multiput(10,0)(0,1){10}{\line(0,1){0.5}}
\put(0,5){\vector(0,1){1}}
\put(10,5){\vector(0,1){1}}
\put(5,0){\vector(1,0){1}}
\put(5,10){\vector(1,0){1}}
\put(5,-1){\makebox(0,0)[ct]{$D$}}
\put(5,11){\makebox(0,0)[cb]{$C$}}
\end{picture}\]\vspace{0mm}
\paragraph{\it Definition:} For $f \in F$ we can define two algebra homomorphisms
$F \rightarrow U$:
\eq
R^+(f)  = \la R, f \ot \id\ra ,\qquad R^-(f) = \la \tilde R, \id \ot f\ra \label{rplus}
\en
\paragraph{\it Yang-Baxter Equation.}
As a consequence of the axioms of a quasitriangular face Hopf algebra $R$
satisfies the Yang-Baxter Equation
\eq
R_{12} R_{13} R_{23} = R_{23} R_{13} R_{12} \quad \in \quad U \ot U \ot U \; .
\en
Contracted 
with $T^A_B \ot T^C_D \ot T^E_F$ this expression yields
a numerical Yang-Baxter equation with the following pictorial representation
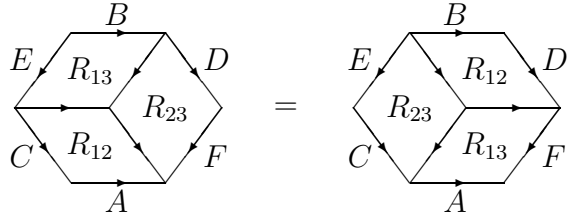
\cite{Bax}:
\begin{center}
\unitlength 0.50mm
\linethickness{0.4pt}
\begin{picture}(145.00,46.00)
\put(0.00,20.00){\line(3,-4){15.00}}
\put(15.00,0.00){\line(1,0){25.00}}
\put(40.00,0.00){\line(3,4){15.00}}
\put(55.00,20.00){\line(-3,4){15.00}}
\put(40.00,40.00){\line(-1,0){25.00}}
\put(15.00,40.00){\line(-3,-4){15.00}}
\put(90.00,20.00){\line(3,-4){15.00}}
\put(105.00,0.00){\line(1,0){25.00}}
\put(130.00,0.00){\line(3,4){15.00}}
\put(145.00,20.00){\line(-3,4){15.00}}
\put(130.00,40.00){\line(-1,0){25.00}}
\put(105.00,40.00){\line(-3,-4){15.00}}
\put(0.00,20.00){\line(1,0){25.00}}
\put(25.00,20.00){\line(3,4){15.00}}
\put(25.00,20.00){\line(3,-4){15.00}}
\put(105.00,40.00){\line(3,-4){15.00}}
\put(120.00,20.00){\line(-3,-4){15.00}}
\put(120.00,20.00){\line(1,0){25.00}}
\put(15.00,40.00){\vector(-3,-4){9.00}}
\put(0.00,20.00){\vector(3,-4){9.00}}
\put(15.00,40.00){\vector(1,0){15.00}}
\put(0.00,20.00){\vector(1,0){15.00}}
\put(15.00,0.00){\vector(1,0){15.00}}
\put(25.00,20.00){\vector(3,-4){9.00}}
\put(40.00,40.00){\vector(-3,-4){9.00}}
\put(40.00,40.00){\vector(3,-4){9.00}}
\put(55.00,20.00){\vector(-3,-4){9.00}}
\put(90.00,20.00){\vector(3,-4){9.00}}
\put(105.00,40.00){\vector(-3,-4){9.00}}
\put(105.00,40.00){\vector(3,-4){9.00}}
\put(120.00,20.00){\vector(-3,-4){9.00}}
\put(130.00,40.00){\vector(3,-4){9.00}}
\put(145.00,20.00){\vector(-3,-4){9.00}}
\put(105.00,40.00){\vector(1,0){15.00}}
\put(120.00,20.00){\vector(1,0){15.00}}
\put(105.00,0.00){\vector(1,0){15.00}}
\put(20.00,30.00){\makebox(0,0)[cc]{$R_{13}$}}
\put(40.00,20.00){\makebox(0,0)[cc]{$R_{23}$}}
\put(20.00,10.00){\makebox(0,0)[cc]{$R_{12}$}}
\put(105.00,20.00){\makebox(0,0)[cc]{$R_{23}$}}
\put(125.00,10.00){\makebox(0,0)[cc]{$R_{13}$}}
\put(125.00,30.00){\makebox(0,0)[cc]{$R_{12}$}}
\put(72.00,20.00){\makebox(0,0)[cc]{$=$}}
\put(27.00,42.00){\makebox(0,0)[cb]{$B$}}
\put(27.00,-2.00){\makebox(0,0)[ct]{$A$}}
\put(5.00,30.00){\makebox(0,0)[rb]{$E$}}
\put(5.00,10.00){\makebox(0,0)[rt]{$C$}}
\put(50.00,10.00){\makebox(0,0)[lt]{$F$}}
\put(50.00,30.00){\makebox(0,0)[lb]{$D$}}
\put(117.00,42.00){\makebox(0,0)[cb]{$B$}}
\put(117.00,-2.00){\makebox(0,0)[ct]{$A$}}
\put(95.00,30.00){\makebox(0,0)[rb]{$E$}}
\put(95.00,10.00){\makebox(0,0)[rt]{$C$}}
\put(140.00,10.00){\makebox(0,0)[lt]{$F$}}
\put(140.00,30.00){\makebox(0,0)[lb]{$D$}}
\end{picture}
\end{center}
\vspace{1ex}
The inner edges are paths that are summed over. 
Moving along the outer edges of the hexagon 
we will later
read of the shifts in the Yang-Baxter equation for 
dynamical $R$-matrices.

So far we have argued heuristically that the Ruijsenaars system naturally leads to 
graphs and face algebras.
The formal relation between face Hopf algebras and oriented graphs
it is established by a lemma of Hayashi 
(Lemma 3.1 of \cite{Hay1}) which says
that any right or left comodule $M$ of a face 
Hopf algebra $F$ decomposes as a linear space
to a direct sum $M=\oplus_{i,j}M_{ij}$ with indices $i,j$ running
over all vertices. So we can naturally associate paths
from 
$j$ to $i$ to any pair of indices such that $M_{ij}\neq \emptyset$. 
So we may speak (and we really do in the paper) of the 
vectors of a comodule or a module as of paths.
As the dual object $U$ to a face Hopf algebra is again a face Hopf algebra
characterized by the same set of vertices \cite{Hay1}, 
the same applies to its 
comodules. It is convenient to choose the orientation of paths
appearing in the decomposition of a comodule of $U$ 
(and hence in the module of $F$)
opposite to the convention that one uses in the case of $F$.

Particularly we have for any matrix corepresentation
$T^A_B$ of $F$ with symbols $A, B$ used to label some 
linear basis in $M$, the linear span $\langle T^A_B \rangle$ of all $T^A_B$ decomposes as 
linear space (bicomodule
of the face Hopf algebra) as a direct sum
\[
\bigoplus_{i,j,k,l}\langle e^ie_jT^A_B e^ke_l \rangle = \langle T^A_B \rangle 
\]
i.e. a sum over paths with fixed starting
and ending vertices. The upper indices $i$ and $k$ fix the beginning and the end
of the path $A$ and the lower indices $j$ and $l$ fix the beginning and the end 
of the path $B$.

Let us assume that
the matrix elements $T^A_B$ of a corepresentation of $F$
act in a module of paths that we shall label by greek characters 
$\alpha$, $\beta$, etc.
The definition of the dual face Hopf algebra implies
that the matrix element $(T^A_B)^{\alpha}_{\beta}$ is nonzero only if 
$\cdot \alpha = \cdot B$, $\cdot A=\alpha \cdot$, $ B\cdot=\cdot \beta$ and
$A\cdot = \beta \cdot $, \ie\ if paths $\alpha\cdot A$ and $B\cdot\beta$ have
common starting and endpoints. This justifies the pictorial 
representation used
in the paper.
It also follows immediately that in the case of a coquasitriangular
Face Hopf algebra 
$
\la \R , T^A_B \ot T^C_D \ra = R^{A C}_{B D}  
\equiv W\Big( {\mbox{\scriptsize$C$} {B \atop A} \mbox{\scriptsize$D$}} \Big)
$
is zero unless $\cdot C = \cdot B$, $B\cdot = \cdot D$,
$C\cdot = \cdot A$ and $A\cdot = D\cdot$.

To make a contact with the Ruijsenaars type of models, 
we have to assume that 
the (coquasitriangular) face Hopf algebra $F$ is 
generated by the matrix elements of some fundamental 
corepresentation of it. We shall postulate the paths
of the corresponding corepresentation to be of length~$1$.
The paths belonging to the $n$-fold tensor product of the 
fundamental corepresentation are then by definition of length~$n$. 
Taking an infinite tensor
product of the fundamental corepresentation we get a graph that corresponds
to the one generated by the shift operators of the related integrable model.

In the next section we are going to formulate a quantum version of the
so-called Main Theorem which gives the solution by factorization of the
Heisenberg equations of motion. For this construction $F$ needs to have a
coquasitriangular structure -- this will  
also fix its algebra structure.

\section{Quantum factorization}

The cocommutative functions in $F$ are of particular interest to us
since they form a set of mutually commutative operators. We shall
pick a Hamiltonian from this set. Cocommutative
means that the result of an application of comultiplication $\Delta$ 
is invariant under exchange of the two resulting factors. The typical
example is a trace of the T-matrix. The following theorem gives for
the case of face Hopf algebras what has become known as the
``Main theorem'' for the solution by factorization of the equations
of motion \cite{Adl,Kost,Sym,RS,Res}. The following theorem 
is a direct generalization of our
previous results for Hopf algebras/Quantum Groups \cite{SJ,PB}:

\begin{thm}[Main theorem for face algebras]
\mbox{ }

\begin{enumerate}
\item[(i)] The set of cocommutative functions, denoted $I$, is a commutative
subalgebra of $F$.
\item[(ii)] The Heisenberg 
equations of motion defined by a Hamiltonian $\ha \in I$
are of Lax form
\eq
i \frac{d T}{dt} = \left[ M^\pm , T \right], \label{laxeqn}
\en
with $M^\pm = 1 \ot \ha - m_\pm \in U_\pm \ot F$, 
$m_\pm = R^\pm(\ha_{(2)}) \ot \ha_{(1)}$; see (\ref{rplus}).
\item[(iii)] Let $g_\pm(t) \in U_\pm \ot F$
be the solutions to the factorization problem
\eq
g_-^{-1}(t) g_+(t) = \exp( i t (m_+ - m_-)) \: \in \: U \ot F ,
\en
then 
\eq
T(t) = g_\pm(t) T(0) g_\pm(t)^{-1} 
\en
solves the Lax equation {\rm (\ref{laxeqn})}; $\:g_\pm(t)$ are given by 
\eq
g_\pm(t) = \exp(-it(1\ot h))\,\exp(it(1\ot h - M^\pm(0))
\en
and are the solutions to the differential equation
\eq
i  \frac{d}{dt}g_\pm(t) =
M^\pm(t) g_\pm(t), \qquad  g_\pm(0) = 1 . 
\en
\end{enumerate}
\end{thm}
\paragraph{\it Proof:}
The proof is similar to the one given in \cite{PB} for factorizable
quasitriangular Hopf algebras. Here we shall only emphasize the points
that are different because we are now dealing with face algebras.
An important relation that we shall use several times in the proof
is
\eq
\Delta(1) T_1 T_2 = T_1 T_2 = T_1 T_2 \Delta(1).
\en
(Note that the generalization is not trivial since
now $\Delta(1) \neq 1 \ot 1$ in general.)
\begin{enumerate}
\item[(i)] Let $f , g \in I \subset F$, then $f g \in I$. Let us show that
$f$ and $g$ commute:
\begin{quote}
$\displaystyle f g = \la T_1 T_2 , f \ot g \ra = \la \Delta(1) T_1 T_2 , f
\ot g \ra = \la \tilde R R T_1 T_2 , f \ot g \ra = \la \tilde R T_2 T_1 R, f
\ot g \ra = \la T_2 T_1 R \tilde R, f \ot g \ra = \la T_2 T_1 \Delta'(1), f
\ot g \ra = \la T_2 T_1, f \ot g \ra = g f. $
\end{quote}
($\tilde R \equiv (S \ot \id)(R)$ can be commuted with $T_2 T_1 R$ in the
fifth step because $f$ and $g$ are both cocommutative.)
\item[(ii)] We need to show that 
$\left[ R^\pm(\ha_{(2)}) \ot \ha_{(1)} , T \right] = 0$. 
This follows from the cocommutativity of $\ha$ and (\ref{rtt}):
\begin{quote}
$\displaystyle \la T_1 R^\pm_{21} T_2 , \ha \ot \id \ra
= \la R^\pm_{21} T_1 T_2 , \ha \ot \id \ra
= \la T_2 T_1 R^\pm_{21} , \ha \ot \id \ra .$
\end{quote}
\item[(iii)] We need to show that $[ m_+ , m_- ] = 0 $; then the proof of 
\cite{PB} applies.\\[1ex]
$\displaystyle m_+ m_- = 
\la T_1 R_{13} T_2 \tilde R_{32} , \ha \ot \ha \ot id\ra
\equiv \la T_1 T_2 R_{13} \tilde R_{32} , \ha \ot \ha \ot \id \ra \\
= \la \tilde R_{12} T_2 T_1 R_{12} R_{13} \tilde R_{32}, 
\ha \ot \ha \ot \id\ra
= \la \tilde R_{12} T_2 T_1 \tilde R_{32} R_{13} R_{12}, 
\ha \ot \ha \ot \id \ra \\
= \la R_{12} \tilde R_{12} T_2 T_1 \tilde R_{32} R_{13} , 
\ha \ot \ha \ot \id \ra
= \la \Delta'(1) T_2 T_1 \tilde R_{32} R_{13} , 
\ha \ot \ha \ot \id \ra \\ 
= \la T_2 \tilde R_{32} T_1 R_{13} , \ha \ot \ha \ot \id \ra
= m_- m_+ . $
\end{enumerate}

\paragraph{\it Remark:} The objects in this theorem
($M^\pm, m_\pm, g_\pm(t), T(t)$)
can be interpreted (a) as elements of
$U \ot F$, (b) as maps $F \rightarrow F$ or (c), 
when a representation of $U$ is
considered, as matrices with $F$-valued
matrix elements.

\section{Dynamical operators}

In the main theorem we dealt with expressions that live in $U \ot F$
(and should be understood as maps from $F$ into itself). In this section
we want to write expressions with respect to one fixed vertex. Like we
mentioned in the introduction we are particularly interested in the action of
the Hamiltonian with respect to a fixed vertex.
Since the Hamiltonian is an element of $F$, we shall initially fix the
vertex in this space; due to the definition of the dual face Hopf algebra $U$
this will also fix a corresponding vertex in that space. 
We shall proceed as follows: We will fix a vertex in the
$T$-matrix with the help of $e^\lambda, e_\lambda \in F$ and
will also introduce the corresponding universal $T(\lambda)$.
Next we will contstruct dynamical $R$-matrices
$R^\pm(\lambda)$ as $R^\pm(T(\lambda))$ -- this is
an algebra homomorphism -- and will give the Yang-Baxter and $RTT$ 
equations with
shifts as an illustration.  Finally we shall plug everything into the main
theorem.

Convention for corepresentation $T$ with respect to a fixed vertex:
$T(\lambda)^A_B$ is zero unless  the range (end) of path $A$ is equal to the
fixed vertex $\lambda$. Such a $T$ will map the vector space spanned by
vectors $v^A$ with $A\cdot = \lambda$ fixed to itself. With the help of the
projection operator $e^\lambda \in F$ we can give the following explicit
expression:
\eq
T(\lambda)^A_B \; = \; T^A_B \, e^\lambda
\quad\sim\quad
\unitlength1mm
\begin{picture}(10,8)(0,4)\small
\put(0,0){\line(1,0){10}}
\put(0,10){\line(1,0){10}}
\multiput(0,0)(0,1){10}{\line(0,1){0.5}}
\multiput(10,0)(0,1){10}{\line(0,1){0.5}}
\put(0,5){\vector(0,1){1}}
\put(10,5){\vector(0,1){1}}
\put(5,0){\vector(1,0){1}}
\put(5,10){\vector(1,0){1}}
\put(10,10){\makebox(0,0)[cc]{$\times$}}
\put(11,10){\makebox(0,0)[lb]{$\lambda$}}
\put(5,-1){\makebox(0,0)[ct]{$B$}}
\put(5,11){\makebox(0,0)[cb]{$A$}}
\end{picture}
\en\vspace{4mm}
The universal $T(\lambda)$ is an abstraction of this an is defined
analogously.
\eq
T_{12}(\lambda) \; = \; T_{12} \,(e^\lambda)_2\quad\sim\quad
\unitlength1mm
\begin{picture}(10,10)(0,4)\small
\put(0,0){\line(1,0){10}}
\put(0,10){\line(1,0){10}}
\multiput(0,0)(0,1){10}{\line(0,1){0.5}}
\multiput(10,0)(0,1){10}{\line(0,1){0.5}}
\put(0,5){\vector(0,1){1}}
\put(10,5){\vector(0,1){1}}
\put(5,0){\vector(1,0){1}}
\put(5,10){\vector(1,0){1}}
\put(10,10){\makebox(0,0)[cc]{$\times$}}
\put(11,10){\makebox(0,0)[lb]{$\lambda$}}
\end{picture}
\en
\vspace{4mm}
\paragraph{\it Coproducts of $\,T$.} Expressions for the coproducts 
$\Delta_1 T(\lambda)$ and $\Delta_2 T(\lambda)$ follow either directly from
the definition of $T(\lambda)$ or can be read of the corresponding pictorial
representations.\\
(i)
Coproduct in $F$-space \cite{Fel}: 
\eq
\Delta_2 T(\lambda) \; = \; T_{12}(\lambda) T_{13}(\lambda - h_2)
\quad
\sim
\quad
\unitlength 1mm
\begin{picture}(15,10)(0,9)\small
\put(10,20){\makebox(0,0)[cc]{$\times$}}
\put(12,20){\makebox(0,0)[lc]{$\lambda$}}
\put(10,10){\makebox(0,0)[cc]{$\times$}}
\put(12,10){\makebox(0,0)[lc]{$\lambda - h_2$}}
\put(0,20){\line(1,0){10}}
\multiput(0,10)(0,1){10}{\line(0,1){0.5}}
\multiput(10,10)(0,1){10}{\line(0,1){0.5}}
\put(0,15){\vector(0,1){1}}
\put(10,15){\vector(0,1){1}}
\put(5,20){\vector(1,0){1}}
\put(0,0){\line(1,0){10}}
\put(0,10){\line(1,0){10}}
\multiput(0,0)(0,1){10}{\line(0,1){0.5}}
\multiput(10,0)(0,1){10}{\line(0,1){0.5}}
\put(0,5){\vector(0,1){1}}
\put(10,5){\vector(0,1){1}}
\put(5,0){\vector(1,0){1}}
\put(5,10){\vector(1,0){1}}
\end{picture}
\en\vspace{6mm}
The shift operator $h$ in $F$-space that appears here is
\eq
h_{(F)} = \sum_{\eta,\mu} (\mu - \eta) \, e^\mu_\eta \; \in \; F,
\en 
where we assume some appropriate (local) embedding of the vertices of the underlying
graph in $\C^n$ so that the difference of vertices makes sense.\\
{\it Proof:} 
$\Delta_2 T_{12} (e^\lambda)_2 
= \sum_\eta T_{12} T_{13} (e_\eta^\lambda)_2 (e^\eta)_3
= \sum_{\eta,\mu} T_{12} (e^\lambda e^\mu)_2 (e_\eta)_2 T_{13} (e^\eta)_3
= \sum_{\eta,\mu} T_{12}(\lambda) (e^\mu_\eta)_2 T_{13}(\lambda + \eta - \mu)
= T_{12}(\lambda) T_{13}(\lambda - h_2)$. In the  second and third step we
used $e^\lambda e^\mu \propto \delta_{\lambda,\mu}$.\\[1mm]
(ii) Coproduct in $U$-space: (gives multiplication
in $F$)
\eq
\Delta_1 T(\lambda) \; = \; T_{13}(\lambda - h_2) T_{23}(\lambda)
\quad \sim \quad
\unitlength 1mm
\begin{picture}(20,10)(0,4)\small
\put(0,0){\line(1,0){10}}
\put(0,10){\line(1,0){10}}
\multiput(0,0)(0,1){10}{\line(0,1){0.5}}
\multiput(10,0)(0,1){10}{\line(0,1){0.5}}
\put(0,5){\vector(0,1){1}}
\put(10,5){\vector(0,1){1}}
\put(5,0){\vector(1,0){1}}
\put(5,10){\vector(1,0){1}}
\put(10,0){\line(1,0){10}}
\put(10,10){\line(1,0){10}}
\multiput(20,0)(0,1){10}{\line(0,1){0.5}}
\put(20,5){\vector(0,1){1}}
\put(15,0){\vector(1,0){1}}
\put(15,10){\vector(1,0){1}}
\put(20,10){\makebox(0,0)[cc]{$\times$}}
\put(20,12){\makebox(0,0)[cb]{\small $\lambda$}}
\put(10,10){\makebox(0,0)[cc]{$\times$}}
\put(10,12){\makebox(0,0)[cb]{\small $\lambda - h_2$}}
\end{picture}\quad
\en \vspace{4mm}
This time we have a shift operator in $U$-space:
\eq
h_{(U)} = \sum_{\eta,\mu} (\mu - \eta) \, E_\mu E^\eta \; \in \; U .
\en
The proof is a little more involved than the one given for the $F$-case above
and uses $(e^\eta)_2 T_{12} = (E^\eta)_1 T_{12}$ and
$T_{12} (e^\mu)_2 = (E_\mu)_1 T_{12}$ which follow from 
$T \tilde T T = T$, $\tilde T T \tilde T = \tilde T$ 
and $T \tilde T = \sum_\xi E^\xi \ot
e^\xi$.
\paragraph{\it Dynamical $R$-matrix.}
Using the fact that
$f \mapsto R^+(f) \equiv \la R , f \ot \id \ra$
is an algebra-homomorphism we define
\eq
R_{12}(\lambda) \equiv  \la R , T_1(\lambda) \ot T_2 \ra
\en
The numerical $R$-matrix is defined analogously:
$R(\lambda)^{AC}_{BD} = \la R, T(\lambda)^A_B \ot T^C_D\ra$.
In the pictorial representation this fixes the one vertex of $R$
that is only
endpoint to paths.

\paragraph{\it Dynamical $RTT$-equation.}
{}From the coproduct $\Delta_1 T$ and $R \Delta(x) = \Delta'(x) R$ for all $x
\in U$ follows \cite{Fel}: 
\eq
R_{12}(\lambda) T_{1}(\lambda - h_2) T_{2}(\lambda)
= T_{2}(\lambda - h_1) T_{1}(\lambda) R_{12}(\lambda - h_3)
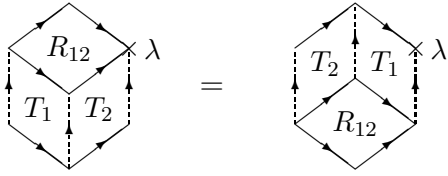
\en
Pictorially:
\[
\unitlength2mm
\begin{picture}(8,6)(0,5)\small
\put(8,8){\makebox(0,0)[cc]{$\times$}}
\put(9,8){\makebox(0,0)[lc]{\small $\lambda$}}
\put(0,3){\line(4,-3){4}}
\put(2,1.5){\vector(4,-3){0.5}}
\put(0,8){\line(4,-3){4}}
\put(2,6.5){\vector(4,-3){0.5}}
\put(0,8){\line(4,3){4}}
\put(2,9.5){\vector(4,3){0.5}}
\put(4,0){\line(4,3){4}}
\put(6,1.5){\vector(4,3){0.5}}
\put(4,5){\line(4,3){4}}
\put(6,6.5){\vector(4,3){0.5}}
\put(4,11){\line(4,-3){4}}
\put(6,9.5){\vector(4,-3){0.5}}
\multiput(0,3)(0,0.5){10}{\line(0,1){0.25}}
\put(0,5.5){\vector(0,1){0.5}}
\multiput(4,0)(0,0.5){10}{\line(0,1){0.25}}
\put(4,2.5){\vector(0,1){0.5}}
\multiput(8,3)(0,0.5){10}{\line(0,1){0.25}}
\put(8,5.5){\vector(0,1){0.5}}
\put(2,4){\makebox(0,0)[cc]{$T_1$}}
\put(6,4){\makebox(0,0)[cc]{$T_2$}}
\put(4,8){\makebox(0,0)[cc]{$R_{12}$}}
\end{picture} \qquad = \qquad
\begin{picture}(8,6)(0,5)\small
\put(8,8){\makebox(0,0)[cc]{$\times$}}
\put(9,8){\makebox(0,0)[lc]{\small $\lambda$}}
\put(0,3){\line(4,-3){4}}
\put(2,1.5){\vector(4,-3){0.5}}
\put(0,8){\line(4,3){4}}
\put(2,9.5){\vector(4,3){0.5}}
\put(4,0){\line(4,3){4}}
\put(6,1.5){\vector(4,3){0.5}}
\put(4,11){\line(4,-3){4}}
\put(6,9.5){\vector(4,-3){0.5}}
\put(0,3){\line(4,3){4}}
\put(2,4.5){\vector(4,3){0.5}}
\put(4,6){\line(4,-3){4}}
\put(6,4.5){\vector(4,-3){0.5}}
\multiput(0,3)(0,0.5){10}{\line(0,1){0.25}}
\put(0,5.5){\vector(0,1){0.5}}
\multiput(4,6)(0,0.5){10}{\line(0,1){0.25}}
\put(4,8.5){\vector(0,1){0.5}}
\multiput(8,3)(0,0.5){10}{\line(0,1){0.25}}
\put(8,5.5){\vector(0,1){0.5}}
\put(2,7){\makebox(0,0)[cc]{$T_2$}}
\put(6,7){\makebox(0,0)[cc]{$T_1$}}
\put(4,3){\makebox(0,0)[cc]{$R_{12}$}}
\end{picture}
\]
\vspace{8mm}
Shifts $h_1$, $h_2$ are in $U$-space, shift $h_3$ is in $F$-space.
Twice contracted with $R$ the dynamical $RTT$-equation yields the dynamical
Yang-Baxter equation:
\paragraph{\it Dynamical Yang-Baxter equation \cite{Ger}} 
\eq
R_{12}(\lambda) R_{13}(\lambda - h_2) R_{23}(\lambda)
= R_{23}(\lambda - h_1) R_{13}(\lambda) R_{12}(\lambda - h_3)
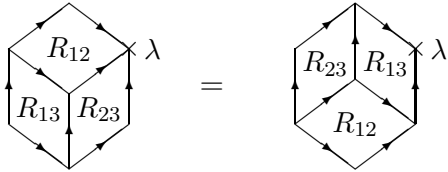
\en
Pictorially:
\[\unitlength2mm
\begin{picture}(8,6)(0,5)\small
\put(8,8){\makebox(0,0)[cc]{$\times$}}
\put(9,8){\makebox(0,0)[lc]{\small $\lambda$}}
\put(0,3){\line(4,-3){4}}
\put(2,1.5){\vector(4,-3){0.5}}
\put(0,8){\line(4,-3){4}}
\put(2,6.5){\vector(4,-3){0.5}}
\put(0,8){\line(4,3){4}}
\put(2,9.5){\vector(4,3){0.5}}
\put(4,0){\line(4,3){4}}
\put(6,1.5){\vector(4,3){0.5}}
\put(4,5){\line(4,3){4}}
\put(6,6.5){\vector(4,3){0.5}}
\put(4,11){\line(4,-3){4}}
\put(6,9.5){\vector(4,-3){0.5}}
\put(0,3){\line(0,1){5}}
\put(0,5.5){\vector(0,1){0.5}}
\put(4,0){\line(0,1){5}}
\put(4,2.5){\vector(0,1){0.5}}
\put(8,3){\line(0,1){5}}
\put(8,5.5){\vector(0,1){0.5}}
\put(2,4){\makebox(0,0)[cc]{$R_{13}$}}
\put(6,4){\makebox(0,0)[cc]{$R_{23}$}}
\put(4,8){\makebox(0,0)[cc]{$R_{12}$}}
\end{picture}
\qquad = \qquad
\begin{picture}(8,6)(0,5)\small
\put(8,8){\makebox(0,0)[cc]{$\times$}}
\put(9,8){\makebox(0,0)[lc]{\small $\lambda$}}
\put(0,3){\line(4,-3){4}}
\put(2,1.5){\vector(4,-3){0.5}}
\put(0,8){\line(4,3){4}}
\put(2,9.5){\vector(4,3){0.5}}
\put(4,0){\line(4,3){4}}
\put(6,1.5){\vector(4,3){0.5}}
\put(4,11){\line(4,-3){4}}
\put(6,9.5){\vector(4,-3){0.5}}
\put(0,3){\line(4,3){4}}
\put(2,4.5){\vector(4,3){0.5}}
\put(4,6){\line(4,-3){4}}
\put(6,4.5){\vector(4,-3){0.5}}
\put(0,3){\line(0,1){5}}
\put(4,6){\line(0,1){5}}
\put(8,3){\line(0,1){5}}
\put(0,5.5){\vector(0,1){0.5}}
\put(4,8.5){\vector(0,1){0.5}}
\put(8,5.5){\vector(0,1){0.5}}
\put(2,7){\makebox(0,0)[cc]{$R_{23}$}}
\put(6,7){\makebox(0,0)[cc]{$R_{13}$}}
\put(4,3){\makebox(0,0)[cc]{$R_{12}$}}
\end{picture}
\]
\vspace{8mm}
\paragraph{\it Hamiltonian and Lax operators in dynamical setting}
\mbox{ }\\
The Hamiltonian $\ha$ should be a cocomutative element of $F$. In the case of the
Ruijsenaars model it can be chosen to be the trace of a $T$-matrix, \ie\ the
$U$-trace of $T$ in an appropriate representation $\rho$: 
$\ha = \mbox{tr}^{(\rho)}_1 T_1$.
This can be written as a sum over vertices $\lambda$ of operators that act in
the respective subspaces corresponding to paths ending in the vertex
$\lambda$:
\eq
\ha = \sum_{Q, \no Q \,{\rm fixed}} T^Q_Q = \sum_\lambda \ha(\lambda)
\en
with
$\ha(\lambda) = \ha e^\lambda = \sum T(\lambda)^Q_Q$.
The pictorial representation of the Hamiltonian is two closed dashed
paths ($F$-space) connected by paths $Q$ of fixed length
that are summed over. In $\ha(\lambda)$ the end of path
$Q$ is fixed. When we look at a representation on Hilbert space the
paths $Q$ with endpoint $\lambda$ that appear in the component
$\ha(\lambda)$ of the Hamiltonian $\ha$ will shift the argument
of a state $\psi(\lambda)$
corresponding to the vertex $\lambda$ to a new vertex corresponding to the
starting point of the path $Q$. 
In the next section we will see in detail
how this construction is applied to the Ruijsenaars system.
For this we will have to take a representation of the face Hopf algebra $F$. We
shall then denote the resulting Lax operator by $L$. The Hamiltonian
will contain a sum 
(coming from the trace) over shift
operators.

For the Lax operators it is convenient to fix two vertices 
$\lambda$ and $\mu$
corresponding
to paths from $\mu$ to $\lambda$ . (We did not need to do this for the
Hamiltonian since it acts only on closed paths.)
The $T$ that appears in the Lax equation (\ref{laxeqn})
becomes
\[T_{12}(\lambda,\mu) = T_{12} (e_\mu e^\lambda)_{2} =  
(E_\lambda)_1 T_{12} (E_\mu)_1 ;\]
it should be taken in some representation of $F$ on the appropriate
Hilbert space. On space $U$ we are interested in a finite-dimensional
representation that is going to give us matrices.
Both is done in the next section and 
we shall call the resulting operator $L(\lambda,\mu)$. 
Similarly we proceed with the other two Lax operators $M^\pm$:
\[
M^\pm(\lambda,\mu) \equiv (E_\lambda \ot 1)M^\pm(E_\mu\ot 1)
= E_\lambda \delta_{\lambda,\mu}\ot \ha
- m_\pm(\lambda,\mu).
\]
(Recall that $E_\lambda E_\mu = E_\lambda \delta_{\lambda,\mu}$.)
If we define $M^\pm_{012} = T_{02} (1 - R^\pm_{01})$ then $M^\pm_{12}
= \la M^\pm_{012} , \ha \ot \id^2 \ra$ and we can write dynamical Lax
equations in an obvious notation as
\begin{eqnarray}
i \frac{d T(\lambda,\mu)}{d t} 
& = & \Big\la M^\pm_{012}(\lambda, \lambda - h_0)
T_{12}(\lambda - h_0,\mu) \\
& & - T_{12}(\lambda, \mu - h_0)
M^\pm_{012}(\mu - h_0,\mu) \, , \, \ha \ot \id^2 \Big\ra \nonumber
\end{eqnarray}
where $h_0 = \sum_{\alpha,\beta} (\alpha - \beta) E_\alpha^\beta$ is the shift
operator in the space contracted by the Hamiltonian. If the Hamiltonian is a 
trace we shall find a sum over shifts.

\paragraph{\it Remark:} There is another possible choice of conventions for
the fixed vertex. We could worked with $T_{12}\{\nu\} = (e_\nu)_2 T_{12}$ 
instead
of $T_{12}(\lambda) = T_{12} (e^\lambda)_2$. This would have fixed the lower
left vertex in the pictorial representation of $T^A_B$ and the vertex that is
\emph{starting}-point for all paths in $R\{\nu\}$. The dynamical equations
would of course look a little different from the ones that we have given.

\section{The Lax Pair}

Here we give as an example the Lax pair for the case of the $N$-particle
quantum Ruijsenaars model.
We may think of the face Hopf algebra $F$ as of the elliptic 
quantum group associated 
to $sl(N)$ introduced by Felder \cite{Fel}. We will continue to use the
the symbol $F$ for it. In that case there is an additional
spectral parameter entering all relations in the same way as it is in the case
ordinary quantum groups.

Let $h$ be the Cartan subalgebra of $sl(N)$ and $h^*$ its dual.
The actual graph related to the elliptic quantum group $F$ is 
$h^*\sim \C^{N-1}$.
This is the huge graph of the remark made in the introduction. 
However it decomposes into a continuous family of disconnected graphs, 
each one isomorphic to $-\eta . \Lambda$, the  
$-\eta\in \C$ multiple of the 
weight lattice $\Lambda$ of $sl(N)$, and we can restrict ourselves to this 
one component for simplicity. Correspondingly the shifts in all formulas
are rescaled by a factor $-\eta$.
The space $h^*\sim \C^{N-1}$
itself will be considered as the orthogonal complement of $\C^{N}=
\oplus_{i=1,...,N}\C\varepsilon_i$, 
$\langle \varepsilon_i, \varepsilon_j \rangle
=\delta_{ij}$ with respect to $\sum_{i=1,...,N}\varepsilon_i$.
We write the orthogonal projection
$\epsilon_i=\varepsilon_i - \frac{1}{N}\sum_k \varepsilon_k$ 
for the generator of $h^*$; $\la \epsilon_i , \epsilon_j \ra 
= \delta_{ij} - 1/N$.
The points of the lattice $\eta . \Lambda$ will be denoted by greek
characters $\lambda$, $\mu$, etc. 

The elliptic quantum group is defined by the matrix elements of the 
``fundamental corepresentation''.
This is described with the help of the paths of length
$1$ in the following way:
Let us
associate a one-dimensional linear space 
$V_{\rho,\lambda} \sim \C \eta.\epsilon_k$ and a corresponding path
of length $1$ to any pair
$\lambda, \rho \in \eta . \Lambda$, such that 
$\rho - \lambda = \eta \epsilon_k$, 
for some $k=1,..., N$. We let $V_{\rho,\lambda}=\emptyset$
for all other pairs of vertices. 
The vector space $V$ of the fundamental corepresentation is formed
by all paths of length $1$
\[
V=\bigoplus_{\lambda,\rho} V_{\rho,\lambda}\quad
\sim
\quad
\bigoplus_{\lambda,\rho}\;\,
\unitlength1mm
\begin{picture}(10,0)(0,-1)\scriptsize
\put(0,0){\line(1,0){10}}
\put(5,0){\vector(1,0){1}}
\put(0,0){\makebox(0,0)[cc]{$\times$}}
\put(10,0){\makebox(0,0)[cc]{$\times$}}
\put(0,1){\makebox(0,0)[cb]{$\rho$}}
\put(10,1){\makebox(0,0)[cb]{$\lambda$}}
\end{picture}
\quad = \quad\bigoplus_{\lambda,i}\quad
\begin{picture}(10,0)(0,-1)\scriptsize
\put(0,0){\line(1,0){10}}
\put(5,0){\vector(1,0){1}}
\put(0,0){\makebox(0,0)[cc]{$\times$}}
\put(10,0){\makebox(0,0)[cc]{$\times$}}
\put(0,1){\makebox(0,0)[cb]{$\lambda + \eta \epsilon_i$}}
\put(10,1){\makebox(0,0)[cb]{$\lambda$}}
\end{picture}
\]
As all spaces $V_{\rho,\lambda}$ are at most one-dimensional, we can
characterize the numerical $R$-matrix $R^{AC}_{BD}$ in the fundamental
corepresentation by just four indices
referring to the vertices of the ``square'' defined by paths 
$A,B,C,D$ in the case of nonzero matrix element $R^{AC}_{BD}$. Let us set
$\cdot B =\cdot C= \nu $, $\cdot D=B\cdot =\mu$, $D\cdot =A\cdot= \lambda$
and $\cdot A= C\cdot =\rho$ and 
also
\[ R^{AC}_{BD}=W\Big( {C {B \atop A} D} \Big)\equiv W\Big( {\nu \atop \rho}\,\,
{\mu\atop \lambda} \Big) \quad\sim\qquad
\unitlength 1.00mm
\linethickness{0.4pt}
\begin{picture}(11.00,5.00)(0,-1.00)
\put(0.00,-5.00){\line(0,1){10.00}}
\put(0.00,5.00){\line(1,0){10.00}}
\put(10.00,5.00){\line(0,-1){10.00}}
\put(10.00,-5.00){\line(-1,0){10.00}}
\put(5.00,5.00){\vector(1,0){1.00}}
\put(5.00,-5.00){\vector(1,0){1.00}}
\put(0.00,0.00){\vector(0,-1){1.00}}
\put(10.00,0.00){\vector(0,-1){1.00}}
\put(-1.00,5.00){\makebox(0,0)[rc]{$\nu$}}
\put(-1.00,-5.00){\makebox(0,0)[rc]{$\rho$}}
\put(11.00,5.00){\makebox(0,0)[lc]{$\mu$}}
\put(11.00,-5.00){\makebox(0,0)[lc]{$\lambda$}}
\end{picture}
\quad.\]
Then the non-zero Boltzmann weights as given by \cite{Jim}
are: ($i\neq j$)
\[ W\Big( {\lambda +2\eta \epsilon_i \atop \lambda +\eta \epsilon_i}\,\,
{\lambda +\eta \epsilon_i\atop \lambda}\Big| u \Big)= 1\quad\sim\quad
\unitlength 1.00mm
\begin{picture}(20.00,0.00)(0,-1.00)\small
\put(0.00,-0.50){\line(1,0){9.5}}
\put(0.00,0.00){\line(1,0){9.5}}
\put(4.00,0.00){\vector(1,0){1.00}}
\put(4.00,-0.50){\vector(1,0){1.00}}
\put(10.00,0.00){\line(1,0){9.5}}
\put(10.00,-0.50){\line(1,0){9.5}}
\put(14.00,0.00){\vector(1,0){1.00}}
\put(14.00,-0.50){\vector(1,0){1.00}}
\put(20.50,-0.25){\makebox(0,0)[lb]{$\lambda$}}
\put(19.50,-0.25){\makebox(0,0)[cc]{$\times$}}
\end{picture}
\quad ,\]
\[ W\Big( {\lambda + \eta (\epsilon_i+\epsilon_j) 
\atop \lambda +\eta \epsilon_i}\,\,
{\lambda +\eta \epsilon_i\atop \lambda} \Big| u \Big)
= \frac{\theta(\eta)\theta(-u
+ \lambda_{ij})}{\theta(u+\eta)\theta(\lambda_{ij})}\quad\sim
\quad
\unitlength 1.00mm
\begin{picture}(11.00,5.00)(0,-1.00)\small
\put(0.00,5.00){\line(0,-1){10.00}}
\put(0.00,-5.00){\line(1,0){10.00}}
\put(10,-4.75){\makebox(0,0)[cc]{$\times$}}
\put(0.50,5.00){\line(0,-1){9.50}}
\put(0.50,-4.50){\line(1,0){9.50}}
\put(0.00,0.50){\vector(0,-1){1.00}}
\put(0.50,0.50){\vector(0,-1){1.00}}
\put(5.50,-4.50){\vector(1,0){1.00}}
\put(5.50,-5.00){\vector(1,0){1.00}}
\put(11.00,-5.00){\makebox(0,0)[lb]{$\lambda$}}
\end{picture}
\quad ,\]
and
\[ W\Big( {\lambda + \eta (\epsilon_i+\epsilon_j) 
\atop \lambda +\eta \epsilon_i}\,\,
{\lambda +\eta \epsilon_j\atop \lambda} \Big| u \Big)
= \frac{\theta(u)\theta(\eta+ 
\lambda_{ij})}{\theta(u+\eta)\theta(\lambda_{ij})} 
\quad\sim\quad
\unitlength 1.00mm
\begin{picture}(11.00,5.00)(0,-1.00)\small
\put(0.00,5.00){\line(0,-1){10.00}}
\put(0.00,-5.00){\line(1,0){10.00}}
\put(10.00,-5.00){\line(0,1){10.00}}
\put(10.00,5.00){\line(-1,0){10.00}}
\put(0.00,0.00){\vector(0,-1){1.00}}
\put(5.00,-5.00){\vector(1,0){1.00}}
\put(10.00,0.00){\vector(0,-1){1.00}}
\put(5.00,5.00){\vector(1,0){1.00}}
\put(11.00,-5.00){\makebox(0,0)[lc]{$\lambda$}}
\put(10,-5){\makebox(0,0)[cc]{$\times$}}
\end{picture}
\quad.\]
Here  
$\lambda_{ij}\equiv \lambda_i - \lambda_j =
\langle\lambda,\epsilon_i-\epsilon_j\rangle$, \,
$\theta(u)$ is the Jacobi theta function
\eq
\theta(u)=\sum_{j\in\Z}e^{\pi i(j+\frac{1}{2})^2\tau 
+ 2\pi i(j+\frac{1}{2})
(u+\frac{1}{2})} , \label{theta}
\en
$u \in \C$ is the spectral parameter and $\tau$ is the 
elliptic modulus parameter.
The matrix element $T^A_B (\lambda|u)$ 
(now depending also on the spectral parameter $u$, which also enters
all previous expressions in the standard way) in the fundamental 
corepresentation is also uniquely determined by the value of the 
vertices that fix the length-$1$ paths $A$ and $B$.
Let us use the following notation for it ($i,j=1,\ldots,N$)
\[
\rule{0mm}{15mm} 
T^A_B (\lambda|u) = \sum_\mu  L^i_j(\mu, \lambda|u) e_\mu
\quad\sim\quad
\unitlength1mm
\begin{picture}(15,6)(-12,4)\scriptsize
\put(0,0){\line(1,0){10}}
\put(0,10){\line(1,0){10}}
\multiput(0,0)(0,1){10}{\line(0,1){0.5}}
\multiput(10,0)(0,1){10}{\line(0,1){0.5}}
\put(0,5){\vector(0,1){1}}
\put(10,5){\vector(0,1){1}}
\put(5,0){\vector(1,0){1}}
\put(5,10){\vector(1,0){1}}
\put(0,0){\makebox(0,0)[cc]{$\times$}}
\put(10,0){\makebox(0,0)[cc]{$\times$}}
\put(0,10){\makebox(0,0)[cc]{$\times$}}
\put(10,10){\makebox(0,0)[cc]{$\times$}}
\put(-1,0){\makebox(0,0)[rt]{$\mu + \eta \epsilon_j$}}
\put(11,0){\makebox(0,0)[lt]{$\mu$}}
\put(-1,10){\makebox(0,0)[rb]{$\lambda + \eta \epsilon_i$}}
\put(11,10){\makebox(0,0)[lb]{$\lambda$}}
\put(5,11){\makebox(0,0)[cb]{\small$A$}}
\put(5,-1){\makebox(0,0)[ct]{\small$B$}}
\end{picture}
\]\vspace{4mm}
where 
$\mu =B\cdot $, $\lambda
=A\cdot $, $\mu +\eta\epsilon_j=\cdot B$, 
$\lambda +\eta\epsilon_i=\cdot A$.

As the lattice $\eta . \Lambda$ that we consider is just one 
connected component
of a continuous family of disconnected graphs, the vertices
$\lambda$, $\mu$, \ldots are allowed to take any values in 
$\C^{(N-1)}$. This will be assumed implicitly
in the rest of this section.
Now we need to specify the appropriate representation of 
$F$ which can be read of from \cite{Has}.
We can characterize it by its path decomposition. 
To any pair of vertices
$\lambda, \mu$ we associate a one-dimensional vector space 
$\tilde V_{\lambda,\mu}\sim \C$ (path from $\mu$ to $\lambda$). 
The representation space $\tilde V$ is then
\[ 
\tilde V=\bigoplus_{\lambda,\mu \in h^*}\tilde V_{\lambda,\mu}.
\]
The matrix element $T^A_B (\lambda) \equiv L^i_j(\lambda,\mu |u)$ 
for fixed $A,B$ and hence also with fixed $i$, $j$, $\lambda$, $\mu$ 
is obviously non-zero
only if restricted to act from $\tilde V_{\lambda, \mu}$ 
to $\tilde V_{\lambda +\eta \epsilon_i, \mu +\eta \epsilon_j}$ 
in which case it acts as multiplication by
\begin{equation}
L^i_j(\lambda,\mu|u)
=\frac{\theta(\frac{c\eta}{N} +u +\lambda_i -\mu_j)}{\theta(u)}
\prod_{k\neq i}
\frac{\theta(\frac{c\eta}{N} +\lambda_{k} 
-\mu_j)}{\theta(\lambda_{k}-\lambda_i)}.\label{L}
\end{equation}
Here we used notation 
$\lambda_i=\langle \lambda , \epsilon_i \rangle$ 
for $\lambda \in h^*$. $c\in \C$ will play the role of coupling constant.

The Hamiltonian is chosen in accordance with Section 3 as $\ha=\sum_P
T^P_P$, i.e. the trace in the ``fundamental'' corepresentation. 
In accordance
with the discussion of the previous sections it is non-zero only when
acting on the diagonal subspace (closed paths)
\[ 
H=\bigoplus_{\lambda \in h^*}H_{\lambda}
\equiv\bigoplus_{\lambda \in h^*}\tilde V_{\lambda, \lambda}
\]
of $\tilde V$. So this is the actual state space of the integrable system
under consideration. The Lax operator $M^\pm$
acts from $\tilde V_{\lambda, \mu}$ 
to $\tilde V_{\lambda +\eta \epsilon_i, \mu +\eta \epsilon_j}$ 
as multiplication by $M^\pm{}^l_k(\lambda, \mu | u , v) =
(1 \ot \ha)^l_k(\lambda, \mu |v) - m_\pm{}^l_k(\lambda, \mu | u , v)$ with
\begin{equation}
(1 \ot \ha)^l_k(\lambda,\mu|v) 
= \delta_{\lambda,\mu} \delta^l_k 
\frac{\theta(\frac{c\eta}{N} + v)}{\theta(v)}
\prod_{j'\neq i}
\frac{\theta(\frac{c\eta}{N} 
+\lambda_{j',i})}{\theta(\lambda_{j',i})} , \label{h}
\end{equation}
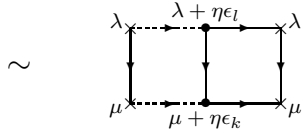
\begin{eqnarray}
\lefteqn{m^+{}^l_k(\lambda,\mu|u , v)  = } \nonumber \\
&& \delta_{\epsilon_i + \epsilon_k , 
\epsilon_j + \epsilon_l}
\delta_{\lambda , \mu + \eta \epsilon_j}
\sum_{i,j} L^i_j(\mu + \eta \epsilon_k , \mu | v)
W\Big( {\lambda + \eta \epsilon_l 
\atop \mu +\eta \epsilon_k}\,\,
{\lambda \atop \mu} \Big| v - u \Big)\\
\rule{0mm}{15mm} && \sim  
\unitlength1mm
\begin{picture}(13,6)(-12,4)\scriptsize
\put(0,0){\line(0,1){10}}
\put(10,0){\line(0,1){10}}
\put(10,0){\line(1,0){10}}
\put(10,10){\line(1,0){10}}
\put(20,0){\line(0,1){10}}
\multiput(0,0)(1,0){10}{\line(1,0){0.5}}
\multiput(0,10)(1,0){10}{\line(1,0){0.5}}
\put(0,5){\vector(0,-1){1}}
\put(10,5){\vector(0,-1){1}}
\put(20,5){\vector(0,-1){1}}
\put(5,10){\vector(1,0){1}}
\put(5,0){\vector(1,0){1}}
\put(15,10){\vector(1,0){1}}
\put(15,0){\vector(1,0){1}}
\put(20,0){\makebox(0,0)[cc]{$\times$}}
\put(20,10){\makebox(0,0)[cc]{$\times$}}
\put(0,0){\makebox(0,0)[cc]{$\times$}}
\put(0,10){\makebox(0,0)[cc]{$\times$}}
\put(10,0){\makebox(0,0)[cc]{$\bullet$}}
\put(10,10){\makebox(0,0)[cc]{$\bullet$}}
\put(21,0){\makebox(0,0)[lt]{$\mu$}}
\put(21,10){\makebox(0,0)[lb]{$\lambda$}}
\put(10,11){\makebox(0,0)[cb]{$\lambda + \eta \epsilon_l$}}
\put(10,-1){\makebox(0,0)[ct]{$\mu + \eta \epsilon_k$}}
\put(-1,0){\makebox(0,0)[rt]{$\mu$}}
\put(-1,10){\makebox(0,0)[rb]{$\lambda$}}
\end{picture} \nonumber
\end{eqnarray}
\mbox{ }\\
$m^-$ is given by a similar formula with the inverse Boltzmann weight.
\paragraph{\it Dynamical Lax Equation}
$$
i\frac{d L^i_k(\lambda, \mu | u)}{d t} =
\sum_{j , \nu} M^\pm{}^i_j(\lambda, \nu | u, v) L^j_k(\nu, \mu |u)
- L^i_j(\lambda, \nu | u) M^\pm{}^j_k(\nu, \mu | u, v)
$$

The Hamiltonian $\ha$ maps 
the component $H_{\lambda}\sim \C.|\lambda \rangle$
into the component 
$H_{\lambda+\eta \epsilon_i}\sim \C.|\lambda +\eta\epsilon_i\rangle$.
Obviously $\tilde V$ can be understood as the 
complex vector space of all functions in $\lambda$, 
$\mu$ as well as $H$ can be understood as the
complex vector space of all functions in $\lambda$. 
In that case
the Hamiltonian $\ha$ given by (\ref{h}) is proportional 
to a difference operator 
in variable $\lambda \in h^*$
\begin{equation}
\ha\propto 
\sum_i t^{(\lambda)}_i\prod_{j\neq i }\frac{\theta(\frac{c\eta}{N} 
+\lambda_{j,i} )}{\theta(\lambda_{j,i})}, \label{Ruijsenaars}
\end{equation}
where $t^{(\lambda)}_i$ has an obvious meaning of the shift operator by 
$-\eta\epsilon_i$ in the variable $\lambda$. 
This is equivalent \cite{Die} to the Ruijsenaars Hamiltonian \cite{Rui}.
It follows from \cite{Has} that in the same way we 
can obtain the higher
order Hamiltonians concerning traces in properly fused 
``fundamental representations".

\subsection*{Acknowledgement}
We would like to thank Pavel Winternitz for constant
interest and support, Koji Hasegawa and Jan Felipe
van Diejen for interesting discussions and Elliott Lieb
for hospitality.

\subsection*{Appendix}

For many reasons it is very convenient to use a formalism based
on the so-called universal $T$. The expressions formally resemble those in
a matrix representation but give nevertheless general face Hopf algebra 
statements. This greatly simplifies notation but also
the interpretation and application of the resulting expressions.
When we are dealing with quantizations of the functions on a group
we need to keep track both of the non-commutativity of the
quantized functions and the residue of the underlying group structure.
Both these structures can be encoded in algebraic relations for the
universal $T$ which easily allows to control
two non-commutative structures. 
In fact $T$ can be regarded as a universal group element.
Universal tensor expressions can \emph{formally} be read in two ways:
Either as ``group'' operations (or rather operations in $U$)
or as the corresponding pull-back maps in the
dual space. This simplifies the heuristics of ``dualizing and reversing
arrows'' and allows us to keep track of the classical limit.
Example: Multiplication in $U$, $x \ot y  \mapsto x y$: The corresponding
pull-back map in the dual space $F$ is the coproduct $\Delta$. Both operations
are summarized in the same universal expression $T_{12} T_{13}$.

Sometimes $T$ can be realized as the canonical element  $U \ot F$, but we
do not need to limit ourselves to these cases an will instead define $T$
as the identity map from $F$ to itself and will
use the same symbol for the identity map $U \rightarrow U$. 

For the application of this identity
map to an element of $F$ we shall nevertheless use the same bracket
notation as we would for a true
canonical element in the finite case, \emph{i.e.} 
$f \equiv \la \id , f \ra = \la T_{12} , f \ot \id\ra$.
This notation is very convenient---like inserting the unity in quantum
mechanics.
The identity map on a product is given by $a b \mapsto a \cdot b$ or
\[ \la T_{12} , a b \ot \id\ra = \la T_{13} T_{23}, a \ot b \ot \id\ra 
= a \cdot b\]
We shall write
$\Delta_1 T_{12} = T_{13} T_{23}$
to express this fact---this is hence the uni\-ver\-sal-T notation for the
multiplication map in $F$ (and the coproduct map in $U$).
The 
coproduct map in $F$ (multiplication map in $U$) is
$T_{12} T_{13}$:
\[f \mapsto \la T_{12} T_{13} , f \ot \id \ot \id\ra = \Delta(f)\]
The antipode map is $\tilde T$: $f \mapsto S(f)$, the
contraction map $x \ot 1$  with $x \in U = F^*$ maps $f \in F$ to
$1 \la x , f \ra$, the
counit map is $1 \ot 1$, \emph{etc.}. 
For face algebras the counit is not an algebra
homomorphism but rather $\Delta 1_{F} = \sum_k e_k \ot e^k$ and 
$\Delta 1_{U} = \sum_k E_k \ot E^k$,
where $\la E_k , f \ra = \epsilon(f e_k)$ and
$\la E^k , f \ra = \epsilon(e^k f)$. Therefore
$\epsilon(a b) = \sum_k \epsilon(a e_k) \epsilon(e^k b)$.
This is one of the face algebra axioms. 
It differs from the corresponding ordinary Hopf algebra axiom
(there: $\epsilon(a b) = \epsilon(a) \epsilon(b)$).
Some important relations involving the coproduct and antipode maps are
\[ \tilde T T = \sum_i E_i \ot e_i, 
\quad T \tilde T = \sum_i E^i \ot e^i,
\quad \tilde T T \tilde T = \tilde T ,
\quad T \tilde T T = T .\]
In the pictorial representation this relations imply that the vertices of the
paths in $U$ and $F$ that form $T$ match to give a closed ``square''.
They also give us two ways to fix the four vertices of $T$;
\[ (1 \ot e^i_k) T (1 \ot e^j_l) \; = \; (E^i_j \ot 1) T (E^k_l \ot 1)
\quad\sim\;
\unitlength1mm
\begin{picture}(15,6)(-12,4)\scriptsize
\put(0,0){\line(1,0){10}}
\put(0,10){\line(1,0){10}}
\multiput(0,0)(0,1){10}{\line(0,1){0.5}}
\multiput(10,0)(0,1){10}{\line(0,1){0.5}}
\put(0,5){\vector(0,1){1}}
\put(10,5){\vector(0,1){1}}
\put(5,0){\vector(1,0){1}}
\put(5,10){\vector(1,0){1}}
\put(0,0){\makebox(0,0)[cc]{$\times$}}
\put(10,0){\makebox(0,0)[cc]{$\times$}}
\put(0,10){\makebox(0,0)[cc]{$\times$}}
\put(10,10){\makebox(0,0)[cc]{$\times$}}
\put(-1,0){\makebox(0,0)[rt]{$k$}}
\put(11,0){\makebox(0,0)[lt]{$l$}}
\put(-1,10){\makebox(0,0)[rb]{$i$}}
\put(11,10){\makebox(0,0)[lb]{$j$}}
\end{picture}
\]\vspace{4mm}
Further useful relations can be obtained from this by summing over some
of the indices and using $\sum_i e^i_j = e_j$, $\sum_j e_j =1$, \emph{etc.} .

\end{document}